\documentclass[amsmath,amssymb,aps]{revtex4-1}
\usepackage{graphicx}
\usepackage{dcolumn}
\usepackage{bm}
\usepackage{gensymb}

\begin{document}

\title{Macroscopically Aligned Carbon Nanotubes as a Refractory Platform for Hyperbolic Thermal Emitters}

\author{Weilu Gao$^{1,\dagger}$}
\author{Chloe F. Doiron$^{1,\dagger}$}
\author{Xinwei Li$^{1}$}
\author{Junichiro Kono$^{1,2,3}$}
\author{Gururaj V. Naik$^{1}$}
\email{guru@rice.edu  $^\dagger$These authors contributed equally.}
\address{$^{1}$Department of Electrical and Computer Engineering, Rice University, Houston, Texas 77005, USA \\
	$^{2}$Department of Physics and Astronomy, Rice University, Houston, Texas 77005, USA \\
	$^{3}$Department of Materials Science and NanoEngineering, Rice University, Houston, Texas 77005, USA}

\begin{abstract}
Refractory nanophotonics, or nanophotonics at high temperatures, can revolutionize many applications, including data storage and waste heat recovery. In particular, nanophotonic devices made from hyperbolic materials are promising due to their nearly infinite photonic density of states (PDOS). However, it is challenging to achieve a prominent PDOS in existing refractory hyperbolic materials, especially in a broad spectral range. Here, we demonstrate that macroscopic films and architectures of aligned carbon nanotubes work as excellent refractory hyperbolic materials. We found that aligned carbon nanotubes are thermally stable up to $1600$\,$\celsius$ and exhibit extreme anisotropy --- metallic in one direction and insulating in the other two directions. Such extreme anisotropy makes this system a hyperbolic material with an exceptionally large PDOS over a broadband spectrum range (longer than $4.3$\,$\mu$m) in the midinfrared, exhibiting strong resonances in deeply sub-wavelength-sized cavities. We observed polarized, spectrally selective thermal emission from aligned carbon nanotube films as well as indefinite cavities of aligned carbon nanotubes with volume as small as $\sim\lambda^3/700$ operating at $700$\,$\celsius$. These experiments suggest that aligned carbon nanotubes possess naturally large PDOS that leads to thermal photon densities enhanced by over two orders of magnitude, making them a promising refractory nanophotonics platform.
\end{abstract}
\maketitle
\pagebreak


Nanophotonics at high temperatures is an emerging technology that enables many new applications. For example, thermophotovoltaic energy conversion, useful for waste heat recovery, can be significantly more efficient with mid-infrared refractory nanophotonic devices~\cite{CouttsFitzgerald1998SA,BasuEtAl2007IJoER,KraemerEtAl2011NM,LenertEtAl2014NN}. Given that the waste heat accounts for 67\% of all energy used in direct energy production in the United States~\cite{energy-llnl} alone, refractory nanophotonics can change the landscape of future energy technology. Additionally, refractory nanophotonics enables novel infrared sources~\cite{GreffetEtAl2002N,SchullerEtAl2009NP,LiuEtAl2011PRL,LiuEtAl2015OME,CostantiniEtAl2015PRA,DyachenkoEtAl2016NC, IlicEtAl2016NN}, ultrahigh density data storage~\cite{ChallenerEtAl2009NP,StipeEtAl2010NP}, and innovative chemical sensing techniques~\cite{InoueEtAl2013APE,LochbaumEtAl2017AP}. In particular, refractory hyperbolic devices possessing huge PDOS hold great promise in sculpting the flow of thermal radiation~\cite{JacobEtAl2012APL,DrachevEtAl2013OE,PoddubnyEtAl2013NP,ShekharEtAl2014NC}, especially in the near-field where the radiative heat transfer can be enhanced beyond Planck's law by several orders of magnitude~\cite{BiehsEtAl2012PRL,GuoEtAl2012APL,GuoJacob2013OE, LiuEtAl2013APL,ZhaoEtAl2017PR,FiorinoEtAl2018NN}. Although natural hyperbolic materials are better in terms of PDOS enhancement than artificial ones, the state-of-the-art refractory hyperbolic material choices suffer from narrow bandwidth and lack of short infrared operation. For example, thermally stable hexagonal form boron nitride (h-BN) and silicon carbide can support hyperbolic dispersions, but only in narrow ranges in 6-12 $\mu$m spectral window~\cite{ShekharEtAl2014NC, CaldwellEtAl2014NC,DaiEtAl2014S,YoxallEtAl2015NP,TamagnoneEtAl2018SA}. Layered graphite possesses hyperbolic dispersions in either ultraviolet range or long infrared wavelength range ($>10$\,$\mu$m). Cuprates and ruthenates~\cite{SunEtAl2014P} supporting broadband hyperbolic dispersions face nanofabrication challenges, and it is not clear whether or not these are refractory materials. Addressing the lack of good refractory hyperbolic materials, here we demonstrate aligned films of single-wall carbon nanotubes (SWCNTs) as an ideal broadband natural hyperbolic refractory material platform for thermal radiation engineering across the mid-infrared.

SWCNTs are one-dimensional materials possessing unique properties, and hence are promising for applications including nanoelectronics and optoelectronics~\cite{AvourisetAl07NN,AvourisetAl08NP}. In particular, the strong quantum confinement results in extremely anisotropic electronic, optical and thermal properties~\cite{DresselhausetAl01Book}, making them exceptional for thermal radiation engineering. To preserve these extraordinary properties at nanoscale in large-scale device applications requires a fabrication technique of producing ensembles of SWCNTs with macroscopic alignment, high packing density, controllable metallicity and chirality, and compatibility with facile nano/micromanufacturing, but it had been a grand challenge of developing successful methods. Recently, we developed a versatile vacuum filtration technique to prepare such aligned SWCNT films~\cite{HeEtAl2016NN}, paving the way for highly anisotropic thermal emitters.

A closely packed bundle of aligned metallic or doped semiconducting SWCNTs can be described by a uniaxial anisotropic medium with effective permittivities $\varepsilon_{\parallel}$ and $\varepsilon_{\perp}$ along the tube axis and in the perpendicular plane, respectively. The conductivity due to free carriers along the nanotubes causes a metallic optical response, leading to Re$(\varepsilon_{\parallel})<0$. In contrast, the nanotubes are insulating in the perpendicular plane, leading to Re$(\varepsilon_{\perp})>0$. This extreme anisotropy leads to a broadband hyperbolic dispersion, which spans a majority of the mid-infrared range, a significant portion of the spectrum of interest for selective thermal emitters. The enhancement of PDOS in hyperbolic materials is limited by the geometry of the material, i.e., the validity limit of the effective medium approximation for high momentum photons (high-\textit{k} modes)~\cite{GuoEtAl2012APL,GuoJacob2013OE}. However, because of the extremely small diameters of SWCNTs ($\sim$1\,nm) compared to the infrared wavelengths, highly aligned and densely packed SWCNT films can support much higher \textit{k} modes, enabling over 1000$\times$ enhancement in PDOS. Because of their hyperbolic dispersions and ultrahigh chemical stability up to 1600\,$\celsius$~\cite{YudasakaEtAl2001NL}, SWCNTs provide a promising platform for mid-infrared refractory nanophotonics and thermal radiation engineering~\cite{MoleskyJacob2015PRB}.

Here, we report hyperbolic thermal emitters emitting spectrally selective and polarized mid-infrared radiation with a 700\,$\celsius$ operating temperature. This novel emitter is based on highly aligned and densely packed SWCNTs prepared through spontaneous alignment that occurs during vacuum filtration~\cite{HeEtAl2016NN}. In comparison to vertically aligned SWCNTs grown by the chemical-vapor-deposition (CVD) method\cite{MurakamiEtAl2004CPL}, our films have a significantly higher packing density, sustain high temperatures, and are robust during nanofabrication. We demonstrate that aligned SWCNT films have a hyperbolic dispersion at wavelengths longer than 4.3\,$\mu$m ($<2335$\,cm$^{-1}$) at 700\,$\celsius$. In direct thermal emission measurements, we observed strongly polarized and narrowband radiation from aligned SWCNT thin films originating from the Berreman modes excited near the epsilon-near-zero (ENZ) frequency. Furthermore, we demonstrate propagating high-\textit{k} photons in the hyperbolic medium, enabling enhanced thermal emission in deep sub-wavelength cavities of volume $\sim\lambda^3/700$, where $\lambda$ is the resonance wavelength. The thermal emission is widely tunable in the mid-infrared range by adjusting the dimensions of the aligned SWCNT structures. 


In order to demonstrate unique optical and thermal radiation properties of the aligned SWCNT structures, we measured their thermal emission and reflectivity at temperatures up to 700\,$\celsius$ using a Fourier transform infrared (FTIR) spectrometer, equipped with a reflective microscope and a heated vacuum stage (see Figure\,1a). See Materials and Methods for more details of the experimental setup. Macroscopically aligned, densely packed SWCNT films were prepared by filtrating a well-dispersed SWCNT suspension through a 2-inch filtration system~\cite{HeEtAl2016NN,KomatsuEtAl2017AFM}. The as-prepared films contained both metallic and semiconducting nanotubes with an average diameter of 1.4\,nm. The obtained films were well aligned and densely packed, as observed in the scanning electron micrograph of Figure\,1b and the cross-sectional transmission electron micrograph in the inset of Figure\,1b. The SWCNT films were transferred onto different substrates using a standard wet transfer technique~\cite{HeEtAl2016NN}. We further fabricated indefinite cavities using electron beam lithography and reactive ion etching. See Materials and Methods and Supplementary Figure\,S1 for the details of the wet transfer and nanofabrication processes, and additional sample characterization data.

\begin{figure}[!h]
	\begin{center}
	\includegraphics[scale=0.6]{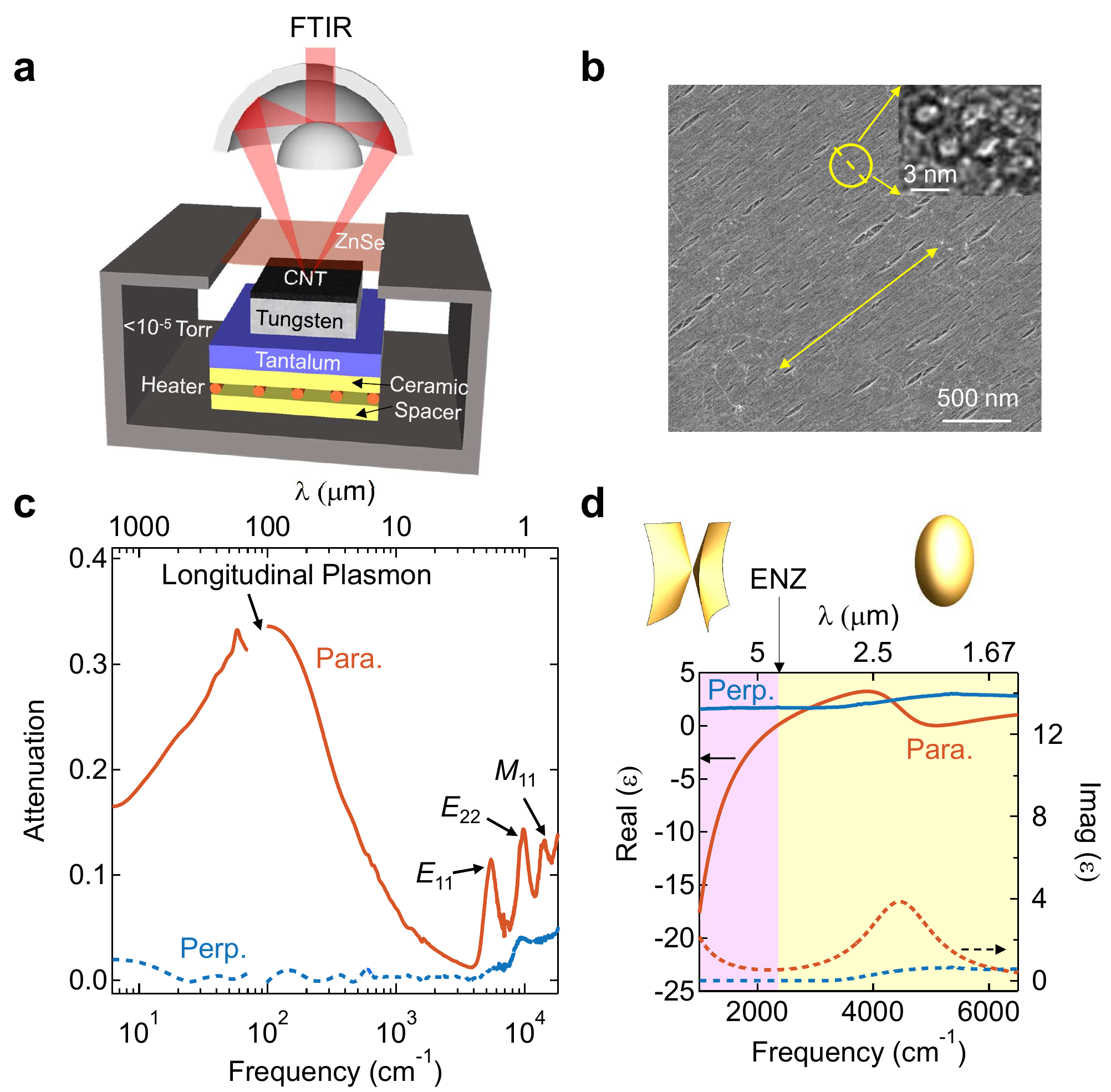}
	\caption{Fabrication and characterization of macroscopically aligned SWCNTs as a mid-infrared hyperbolic material. (a)~Schematic diagram of the experimental setup for thermal emission and reflectivity measurements at temperatures up to 700\,$\celsius$. Samples were heated using a PID-controlled resistive heater surrounded by ceramic spacers under high vacuum $<$\,10$^{-5}$\,Torr. A tantalum film was used to support samples and block the direct thermal emission from the heater. A zinc selenide (ZnSe) window provides optical access for the light collection and analysis using a microscope and a Fourier transform infrared (FTIR) spectrometer, respectively. (b)~A scanning electron micrograph and a transmission electron micrograph (inset), demonstrating a perfect alignment and high packing density. (c)~Polarization attenuation spectra in a wide frequency range, from the THz/far-infrared range to the visible range, at room temperature. (d)~The dielectric constants parallel and perpendicular to the SWCNT alignment direction, with an ENZ frequency in the mid-infrared range.}
     \end{center}
\end{figure}

A room-temperature optical absorption spectrum for a densely packed aligned SWCNT film in a broad spectral range is shown in Figure\,1c, from terahertz (THz) to visible, obtained using THz time-domain spectroscopy, FTIR, and visible-near-IR spectroscopy. The optical absorption is strongly polarization dependent.  When the polarization is perpendicular to the tube axis, the absorption is small and featureless. On the other hand, a prominent absorption peak at $\sim100$\,cm$^{-1}$ is observed for the parallel polarization case arising from the longitudinal plasmon resonance~\cite{ZhangetAl13NL} in finite-length SWCNTs. Further, absorption peaks in the near-infrared and visible ranges, arising from the first two interband transitions for semiconducting tubes ($E_{11}$ and $E_{22}$) and the first interband transition for metallic tubes ($M_{11}$) are clearly observed in the parallel polarization case. 

Wavelength-dependent dielectric constants of the SWCNT films in both directions ($\varepsilon_\parallel$ and $\varepsilon_\perp$) were extracted from reflectance measurements at room temperature and 700\,$\celsius$. A combination of Drude and Lorentz oscillators~\cite{ZhangetAl13NL} were used to retrieve the dielectric function in the parallel direction, and a point-wise retrieval~\cite{DentonEtAl1972JPDAP} was employed for the perpendicular case (see SI for more information). At room temperature, the aligned SWCNT film have a hyperbolic dispersion below 3300\,cm$^{-1}$ ($>3$\,$\mu$m). The room temperature dielectric constants are reported in Supplementary Figure\,S2c. At 700\,$\celsius$, the extracted dielectric functions are very similar to the room temperature measurements except for increased Drude damping and a slight red-shift in the plasma frequency at 700\,$\celsius$. The red-shift in the plasma frequency can be attributed to high-temperature-induced dedoping~\cite{WuetAl04Science}. However, the optical properties of aligned SWCNTs remained stable under vacuum after dedoping during the first heating cycle. Reflectivity measurements during heating cycles are discussed in Supplementary Note 1 and Figure\,S2. 

Figure\,1d shows the real and imaginary parts of the extracted ordinary and extraordinary dielectric constants in both directions at 700\,$\celsius$. The permittivity perpendicular to the tube axis shows a low-loss dielectric behavior over the whole spectral range (Re($\varepsilon_{\perp}$)\,$>$\,0, Im($\varepsilon_\perp$)\,$\approx$\,0), whereas the permittivity parallel to the tube axis is metallic (Re($\varepsilon_{\parallel})<0$) in the mid-infrared with an ENZ frequency of about 2335\,cm$^{-1}$ (4.3\,$\mu$m). While the real permittivities have different signs in different directions for frequencies smaller than the ENZ frequency, they are both positive for higher frequencies. The extreme anisotropy for frequencies smaller than the ENZ frequency results in a hyperbolic dispersion or hyperboloid isofrequency surface, an open surface with an unbounded surface area and hence an unbounded PDOS~\cite{JacobEtAl2012APL}. The nearly infinite PDOS of hyperbolic materials arising from the allowed propagation of extremely high-momentum (high-\textit{k}) waves has many consequences in the context of thermal radiation, including super-Planckian thermal emission, significantly enhanced near-field radiative heat transfer, and thermal conductivity beyond the Stephan-Boltzmann law~\cite{GuoEtAl2012APL,NarimanovSmolyaninov2012}. The aligned SWCNT film supports a hyperbolic dispersion below 2335\,cm$^{-1}$ ($>4.3$\,$\mu$m) at 700\,$\celsius$, covering a significant portion of the region of interest for selective thermal emitters. 

To demonstrate the hyperbolic behavior of aligned SWCNT films, we first characterized the thermal emission from continuous films. The frequency at which the dispersion goes from ellipsoidal to hyperbolic is the ENZ frequency, where we would expect the maximum thermal emission from continuous films~\cite{GuoEtAl2012APL}. Thermal emission measurements of continuous films of aligned SWCNTs on tungsten substrates at 700\,$\celsius$ are shown in Figure\,2. Figure\,2a shows polarization-dependent thermal emission from two different SWCNT films with thicknesses of 188\,nm and 1.75\,$\mu$m, respectively. All emission spectra are normalized with respect to the relatively flat emission spectrum of the tungsten substrate. In the case of the thinner film (188\,nm), we clearly observe a prominent emission peak denoted as $\omega_\text{ENZ}$ in the parallel polarization, whereas no such peak is observed in the perpendicular polarization. The $\omega_\text{ENZ}$ peak originates from the impedance matching condition in the multilayer structure, leading to the excitation of Berreman modes~\cite{HarbeckeEtAl1985APA}. Since no such phenomenon occurs in the perpendicular polarization, we do not see any enhanced thermal emission. The polarization dependence of the thermal emission from the 188\,nm-thick film is shown in Figure\,2b. The experimental data can be fit with a $\cos^2\theta$ curve very well, which is also consistent with the polarization-dependent absorptivity in aligned SWCNTs~\cite{MurakamiEtAl2005PRL}. A contrast over three in emission intensity is observed between the two orthogonal polarizations.  

\begin{figure}[!h]
    \begin{center}
	\includegraphics[scale=0.6]{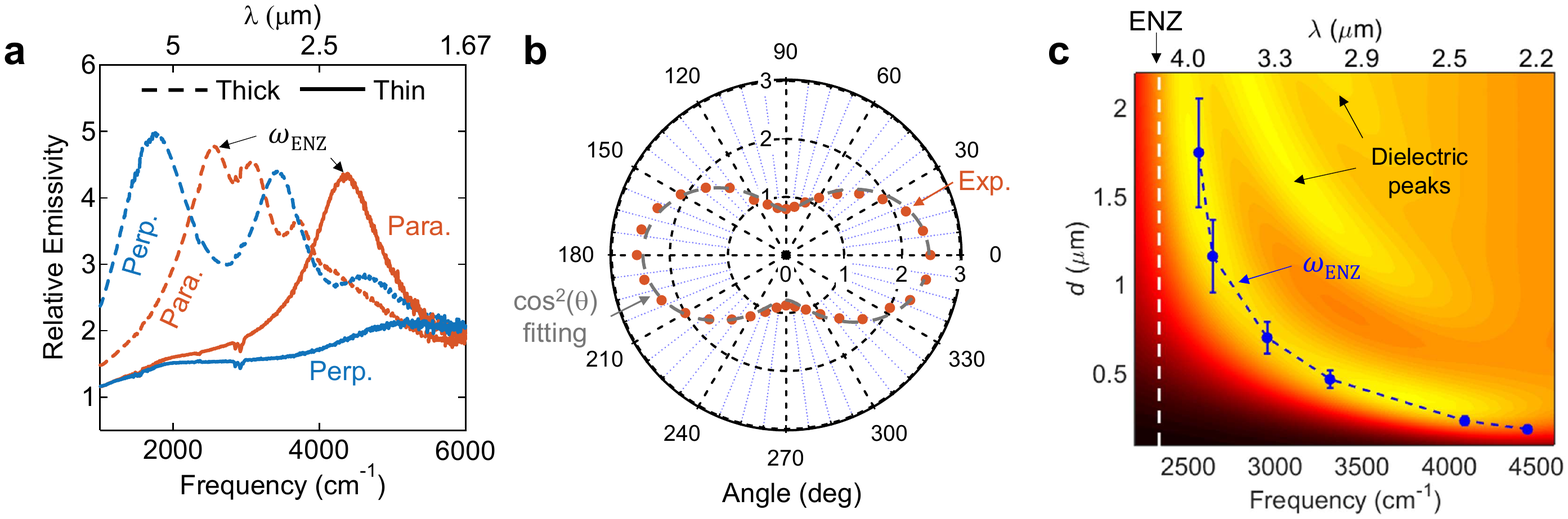}
	\caption{Thermal emission from continuous films of aligned SWCNTs. (a)~Relative emissivity of a 188-nm thick film and a 1.75-$\mu$m thick film on tungsten substrates, parallel and perpendicular to the SWCNT alignment direction, respectively. The measurement was performed at 700\,$\celsius$. The emission spectra were normalized to the spectrum from a bare tungsten substrate, which is featureless. (b)~Polarization-dependent integrated emission signal over frequency for the 188-nm thick film in polar coordinates. A $\text{cos}^2(\theta)$ fitting matches the experimental data well. (c)~The calculated emission spectra for various film thicknesses $d$. The experimental peak position of $\omega_\text{ENZ}$ is overlay as the blue dots connected by a dashed line. A clear asymptomatic approach as thickness increases is observed. As film thickness increases, dielectric peaks also emerge.}
	\end{center}
\end{figure}

Figure\,2c shows calculated emission spectra for various film thicknesses based on the transfer matrix method (hot colormap mapping), and the $\omega_\text{ENZ}$ peak seen in experiments (blue dots connected by a dashed line). As the SWCNT film thickness increases, $\omega_\text{ENZ}$ red-shifts to asymptotically approach the ENZ frequency. Thicker films also support Fabry-Perot resonances, leading to several dielectric peaks in both parallel and perpendicular polarizations, as seen in Figures\,2a and 2c. Thicker samples were fabricated via a manual stacking technique developed recently (see Materials and Methods for more details)~\cite{KomatsuEtAl2017AFM}. The experimental $\omega_\text{ENZ}$ peak matches well with calculations. We clearly observe that $\omega_\text{ENZ}$ approaches the ENZ frequency (white dashed line) when the film thickness increases, as shown in Figure\,2c. The increasing error bars for thicker films are from the cumulative thickness uncertainty in the transfer process. Note that the influence of fabrication imperfections, including imperfect conformal transfer and small misalignment between layers, is expected to be negligible on thermal emission because the standard deviation in the angle of alignment of SWCNT films was found out to be less than two degrees~\cite{HeEtAl2016NN}. The stacking technique can indeed preserve alignment between layers very well and stack films consistently~\cite{KomatsuEtAl2017AFM}.

Extremely large PDOS for hyperbolic materials implies that the medium can support a significantly larger number of thermal photons per unit volume than a blackbody. In other words, a hyperbolic medium can support the same number of thermal photons as a blackbody in a much smaller volume. However, not all of these thermal photons radiate out to far-field because of momentum mismatch. Nanostructuring hyperbolic materials can resonantly outcouple some of these photons and provide information about their momentum. Thus, observing resonances in far-field thermal emission from sub-wavelength cavities of hyperbolic or indefinite materials allows probing the momentum of high-$k$ thermal photons supported by the medium~\cite{YaoEtAl2011PNASU,YangEtAl2012NP}. Figure\,3 summarizes finite-difference-time-domain (FDTD) simulation results of SWCNT indefinite cavities.  We considered a square lattice of SWCNT indefinite cavities on a tungsten substrate with a 400-nm-thick Al$_2$O$_3$ spacer as shown in the schematic of Figure\,3a. The 400-nm-thick Al$_2$O$_3$ spacer reduces any plasmonic interaction between the cavities and the tungsten substrate. We included a 50-nm-thick SiO$_2$ layer on top of SWCNTs, which served as an etch mask during fabrication. The array period and the thickness of the SWCNT layer were fixed at 2.5\,$\mu$m and 500 nm, respectively. The lengths of the cavity in directions parallel and perpendicular to the tube axis are denoted by $L_\parallel$ and $L_\perp$, respectively. Invoking Kirchoff's law, we calculated the emissivity spectra for various $L_\parallel$ and $L_\perp$ combinations. Figure\,3b shows the calculated emissivity relative to the tungsten substrate as a function of frequency for indefinite cavities with various $L_\parallel$ and constant $L_\perp$ at 1.05\,$\mu$m. An emissivity peak is observed only in polarization parallel to the tube axis and originates from the hyperbolic resonance in the deep sub-wavelength cavities. The resonance red-shifts with increasing $L_\parallel$ as expected in any plasmonic or photonic structures. 

\begin{figure}[!h]
	\begin{center}
    \includegraphics[scale=0.6]{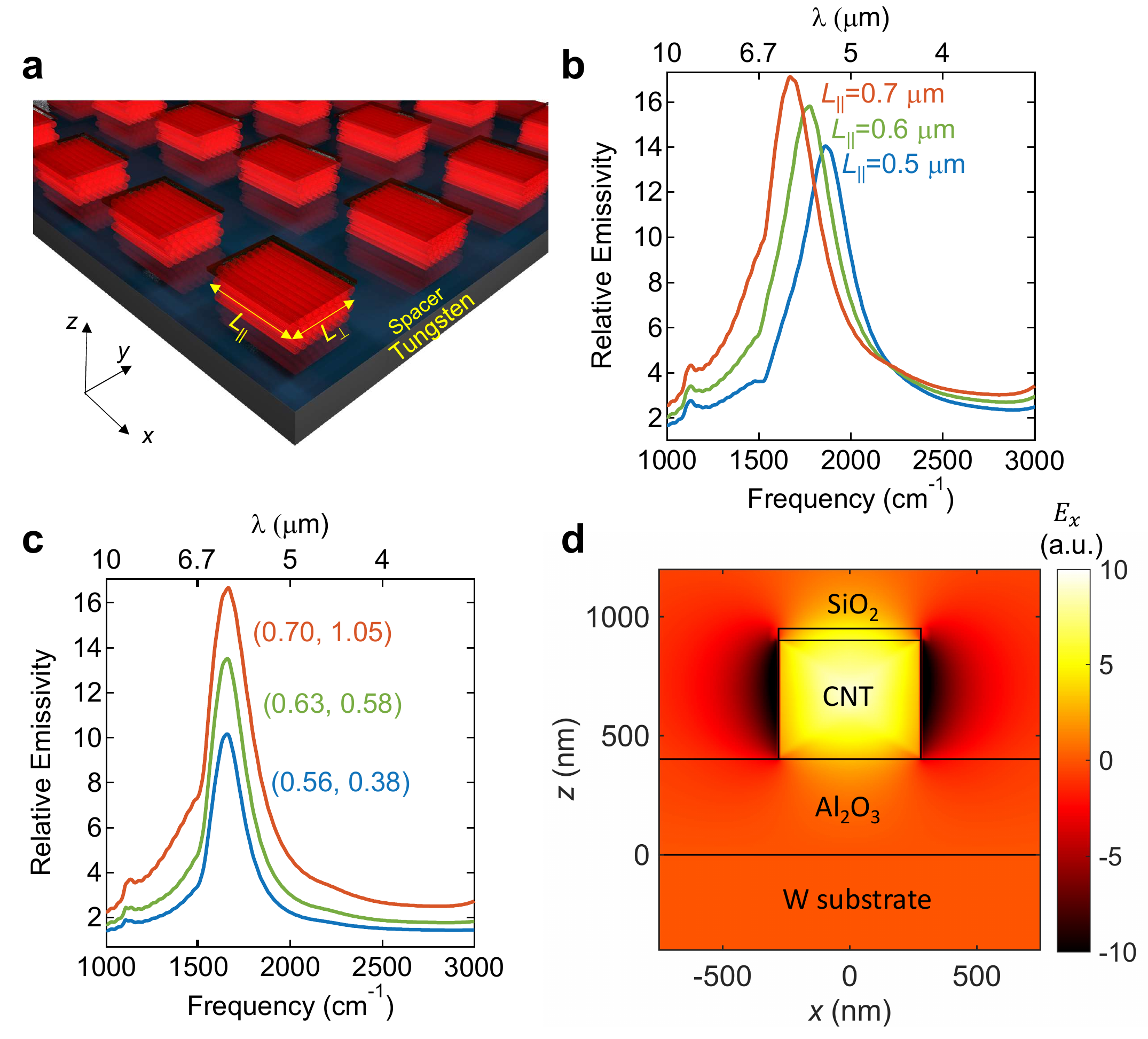}
	\caption{FDTD simulations on an array of SWCNT indefinite cavities. (a)~Schematic diagram of a square lattice of indefinite cavities optimized for selective thermal emission. A spacer is included to suppress the plasmonic interaction with metallic substrates. The dimension along (perpendicular to) the tube axis is $L_\parallel$ ($L_\perp$), and the film thickness is $d$. All quantities are in $\mu$m. (b)~FDTD calculations of relative emissivity for three cavities with tunable $L_\parallel$ and fixed $L_\perp=1.05\,\mu$m. (c)~FDTD calculations of relative emissivity for three cavities tuned along the isofrequency contour. (d)~The electric field ($E_\text{x}$) profile of the (0.56, 0.38) cavity at the peak emissivity.}
    \end{center}
\end{figure}

However, indefinite cavities differ from conventional plasmonic or photonic cavities in that their resonance scales anomalously with $L_\parallel$ and $L_\perp$~\cite{YaoEtAl2011PNASU,YangEtAl2012NP} as demonstrated in Figure\,3c. Here, we designed three cavities with different combinations of $L_\parallel$ and $L_\perp$ in such a way that the emissivity peak occurs at the same frequency. The dimensions of the three cavities chosen in Figure\,3c are indicated in the parentheses as ($L_{\parallel}$, $L_{\perp}$) in $\mu$m. Note that $L_{\parallel}$, $L_{\perp}$ scale the same way unlike in conventional cavities to keep the resonance at the same position. Also, all of these cavities are deeply subwavelength and support the same TM$_{11}$ resonance. Figure\,3d shows the resonant electric field ($E_x$) distribution for the smallest cavity (see Supplementary Figure\,S3 for more data). Unlike plasmonic resonators, the field inside an indefinite cavity is enhanced similar to a photonic resonance. Since the optical losses that lead to thermal radiation are present primarily inside the cavity, the field enhancement effectively leads to enhanced thermal radiation. The divergence of the Poynting vector presented in Figures\,S3a-c shows the location where thermal emission originates.  

We fabricated an array of indefinite cavities; false-color scanning electron micrographs are shown in Figure\,4a. The inset shows the SiO$_2$ mask, 500-nm thick patterned CNT cavities, and the 400-nm thick Al$_2$O$_3$ film on the tungsten substrate. Figure\,4b shows measured relative thermal emission spectra from indefinite cavities with a fixed $L_\perp$ of 1.5\,$\mu$m and three different values of $L_\parallel$. The emissivity peaks red-shift with increasing $L_\parallel$, in qualitative agreement with the simulations results shown in Figure\,3b. Increasing $L_\perp$ while fixing $L_\parallel$ results in a blue-shift; see Supplementary Note 2 and Figure\,S4. 

\begin{figure}[!h]
    \begin{center}
	\includegraphics[scale=0.6]{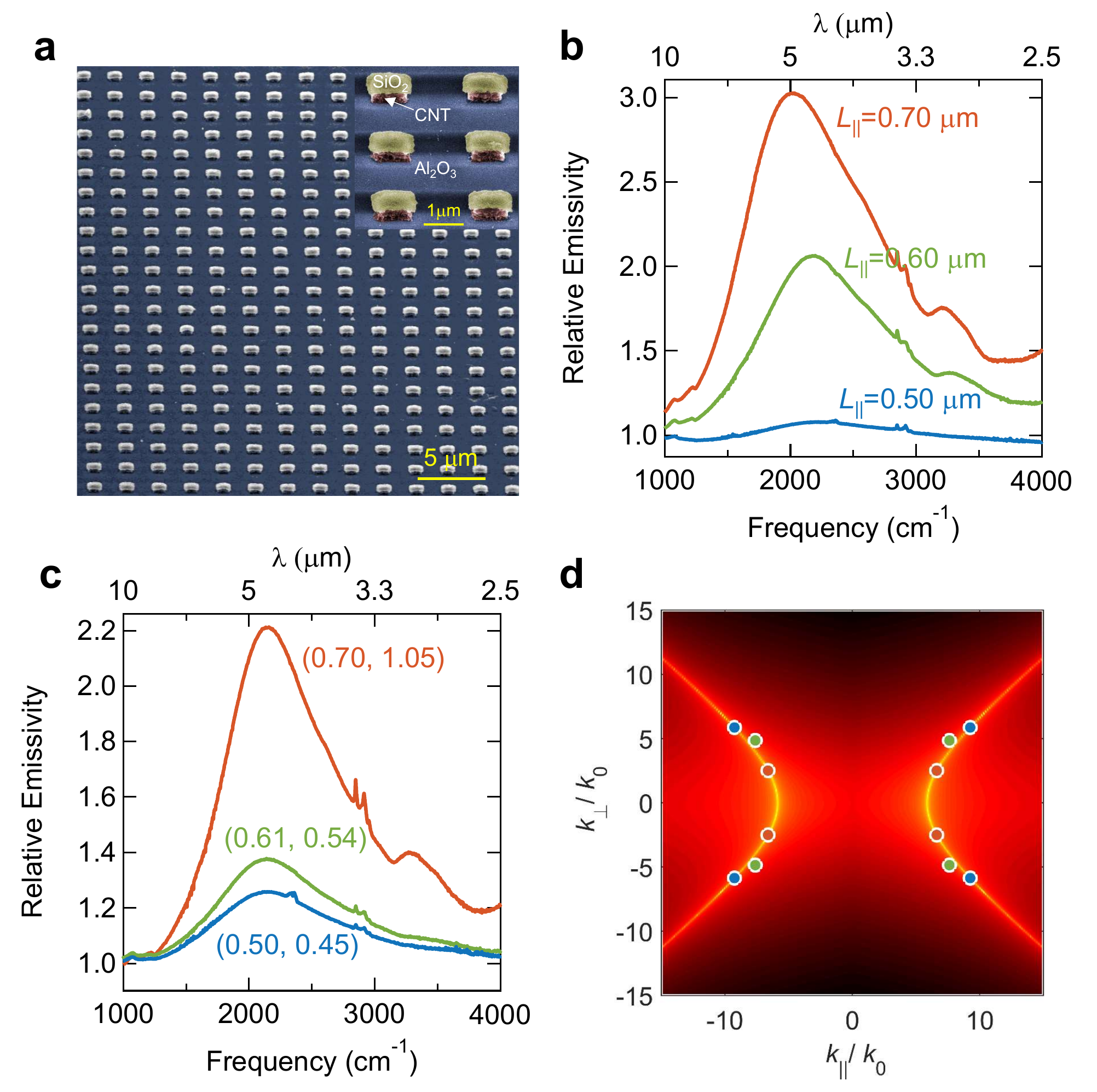}
	\caption{Thermal emission from SWCNT indefinite cavities. (a)~Representative false-color scanning electron micrographs of fabricated SWCNT indefinite cavities. The inset describes a detailed structure of fabricated cavities. (b)~Experimentally measured relative emissivity for three cavities with tunable $L_\parallel$ and fixed $L_\perp=1.5\,\mu$m.  (c)~Experimentally measured relative emissivity of three cavities with different combinations of dimensions in $\mu$m. All cavities show the same emission peak position at 2140 cm$^{-1}$ (4.7\,$\mu$m). (d)~The extracted hyperbolic dispersion based on the dielectric constants at 2140 cm$^{-1}$ from Figure\,1d. $k_\parallel$ is the wavevector component along the tube alignment direction ($x$-axis), $k_\perp$ is that perpendicular to the alignment direction in the film plane ($y$-axis), and $k_0$ is the wavevector in vacuum.}
	\end{center}
\end{figure}

By simultaneously changing $L_\perp$ and $L_\parallel$, we can maintain the resonance frequency at a single frequency, as shown in Figure\,4c. For all three different ($L_{\parallel}$, $L_{\perp}$) combinations, the cavities emit resonantly at 2140\,cm$^{-1}$ (4.7\,$\mu$m). This emission peak is slightly blue-shifted from the simulations and is also broader largely due to fabrication imperfections. By tuning the dimensions of the cavities so that the resonances trace an isofrequency contour, the dispersion of the SWCNT layer can be determined at this frequency. Figure\,4d plots the experimentally measured dispersion points on the isofrequency contour of the SWCNT film calculated from the previously measured dielectric constants of the SWCNT film at 2140\,cm$^{-1}$. Here, $k_\parallel$ is the wavevector component along the tube alignment direction ($x$-axis), $k_\perp$ is that perpendicular to the alignment direction in the film plane ($y$-axis), and $k_0$ is the wavenumber in vacuum. The measured dispersion agrees well with the calculations, confirming the existence of a hyperbolic dispersion behind the observed thermal emission behaviors. Further, the observation of resonances in deep sub-wavelength sized cavities proves that high-\textit{k} waves supported in the SWCNT hyperbolic medium lead to significantly high PDOS or density of thermal photons. The smallest cavity in which we observed a resonance had a volume of $\sim\lambda^3/700$, indicating at least a 100$\times$ enhancement of PDOS in SWCNT hyperbolic thermal emitters.

We demonstrated that aligned SWCNTs make an excellent material platform for refractory nanophotonics in the mid-infrared. Our aligned SWCNT devices showed broadly tunable, polarized, and spectrally selective hyperbolic thermal emitters operating at 700\,$\celsius$. While thin films of aligned SWCNTs exhibited enhanced thermal emission near their ENZ frequency due to the excitation of Berreman modes, nanopatterned films showed geometry-tunable spectrally-selective thermal emission arising from indefinite cavity resonances. These resonances in deep sub-wavelength cavities allowed direct measurements of hyperbolic dispersions in SWCNT films, proving the existence of propagating high-\textit{k} waves and significantly large PDOS. Photonic-like resonances in cavities of volume as small as $\lambda^3/700$ showed that the density of thermal photons in the SWCNT hyperbolic medium is greater than that in a blackbody by at least 100$\times$.

\section{Methods}
\subsection{Thermal emission and reflectivity experimental setup} A Thermo Fisher Scientific Fourier transform infrared (FTIR) spectrometer, equipped with a microscope and a customized high-vacuum PID-controlled heating element, was used to measure the thermal emission and reflectivity of aligned SWCNT structures. The microscope has a reflective infrared objective, with NA\,=\,0.24 (corresponding to a collection angle $\sim14^\circ$) and aperture size 300\,$\mu$m $\times$ 300\,$\mu$m. A zinc selenide (ZnSe) window with anti-reflection coating was mounted on the heating stage for the light collection in the whole mid-infrared region, from 1000 cm$^{-1}$ to 7000 cm$^{-1}$. The heating element consists of a resistive heater surrounded by ceramic spacers. A tantalum film was used to support samples and block the direct thermal emission from the heater. The heating element was kept at target temperatures for at least 30 minutes to reach thermal equilibrium before any measurements. All experiments were performed under high vacuum $<$\,10$^{-5}$\,Torr.

\subsection{SWNCT sample preparation and indefinite cavity fabrication} Arc-discharge P2-SWCNTs with an average diameter 1.4\,nm were purchased from Carbon Solutions, Inc. 20 mL 0.5\% (wt./vol.) sodium deoxycholate (DOC, Sigma-Aldrich) was used to disperse 8\,mg P2-SWCNTs. A tip sonicator was used to homogenize the suspension for 45 minutes, and the obtained suspension was ultracentrifuged for 1.5 hours at 38000 rpm to remove undispersed large bundles and impurities. Well-dispersed supernatant was collected, and then diluted to reduce the surfactant concentration below the critical michelle concentration (CMC) of DOC, which is considered to be a necessary condition for spontaneous alignment. The diluted suspension was then poured into a 2-inch vacuum filtration system to have a uniform, aligned, and densely packed SWCNT film. The obtained film can be transfered to various substrates, by dissolving the filter membrane in chloroform and then rinsing the sample in acetone~\cite{HeEtAl2016NN}. Furthermore, multiple thin aligned films were stacked to have a thicker film by manually transferring several pieces onto the substrate one by one, while preserving the alignment direction~\cite{KomatsuEtAl2017AFM}. 

A 400-nm thick film of Al$_2$O$_3$ was deposited onto a tungsten substrate in an reaction of trimethylaluminum and water using an atomic layer deposition system. A thick film of aligned SWCNTs was transferred onto the Al$_2$O$_3$ film using the aforementioned stacking process. A 50-nm thick film of SiO$_2$ was deposited onto the stacked SWCNT film as a hard mask, using an ultrahigh vacuum DC sputtering system. This hard mask aims to increase the dry etching selectivity for SWCNT patterning. We used standard electron beam lithography to define large-area patterns. A SU-8 positive lithography resist was deposited, following the coating of a thin layer of Omnicoat for easy lift-off. The patterns were first transferred onto the SiO$_2$ hard mask using a sulfur hexafluoride (SF$_6$) based reactive ion etching (RIE), and the SU-8 resist was removed using PG remover. Finally, a pure oxygen RIE was performed to transfer the patterns onto SWCNTs to form indefinite cavities.

\section{Supplementary Information}
\setcounter{figure}{0}
\makeatletter 
\renewcommand{\thefigure}{S\@arabic\c@figure}
\makeatother
\subsection{Reflectance spectra fitting and stability of aligned SWCNTs at 700\,$^\circ $C}
A 110-nm thick film of aligned SWCNTs was transferred onto a CaF$_{2}$ substrate for reflectance measurements. We used a Bruker Multimode 8 atomic force microscope (AFM) to determine the thickness of CNT films. Figure\,S1 demonstrates an AFM image (Figure\,S1a) of a single-layer film with thickness 42\,nm\,$\pm$\,4\,nm (Figure\,S1b). 

\begin{figure}[!h]
	\begin{center}
		\includegraphics[scale=0.7]{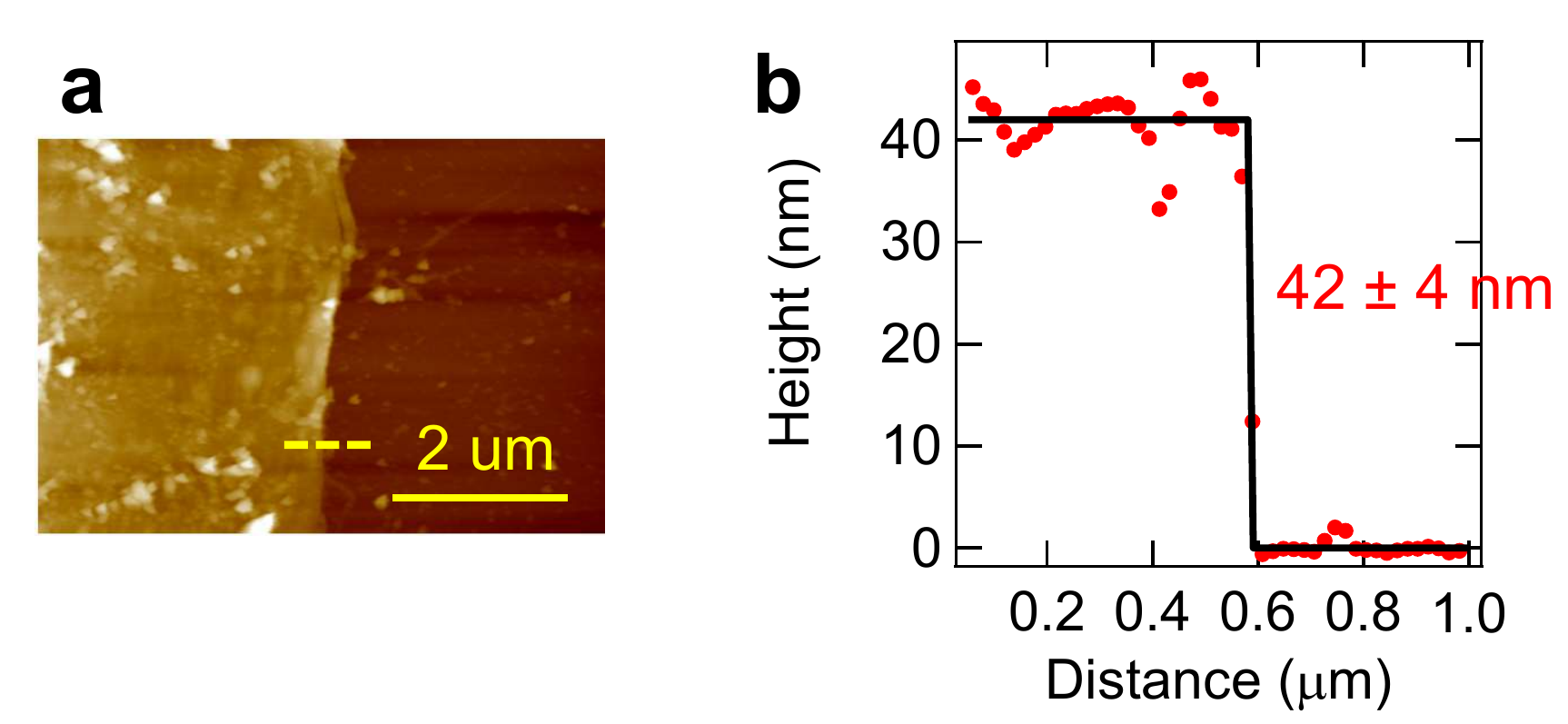}
		\caption{(a)~The atomic force microscope image of an aligned film on a SiO$_2$/Si substrate, and (b)~corresponding height profile across the yellow dashed line in (a).}
	\end{center}
\end{figure}

The reflected signal from SWCNTs on a CaF$_{2}$ substrate was normalized to the reflected signal from the CaF$_{2}$ substrate.  Along the tube alignment direction, the dielectric constant is modeled as
\begin{equation}
\varepsilon=\varepsilon_{\infty}+\varepsilon_{\text{Drude}}+\varepsilon_{\text{THz}}+\varepsilon_\text{{inter}}
\end{equation}
where $\varepsilon_{\text{Drude}}$ captures the response from intertube transport and percolating channels in the macroscopic scale, $\varepsilon_{\text{THz}}$ characterizes the response from the confined collective motion in the tube-length level, and $\varepsilon_\text{{inter}}$ describes the first exciton peak. $\varepsilon_{\text{Drude}}$ has a Drude response as
\begin{equation}
\varepsilon_{\text{Drude}}(\omega)=-\frac{\omega_{\text{Drude}}^2}{\omega^2+i\gamma_{\text{Drude}}\omega}\ ,
\end{equation}	
while $\varepsilon_{\text{THz}}$ and $\varepsilon_\text{{inter}}$ have Lorentzian shapes as
\begin{equation}
\varepsilon_{\text{THz}}(\omega)=\frac{\omega_{\text{p,THz}}^2}{\omega_{\text{0,THz}}^2-\omega^2+i\gamma_{\text{THz}}\omega}
\end{equation}	
\begin{equation}
\varepsilon_{\text{inter}}(\omega)=\frac{\omega_{\text{p,inter}}^2}{\omega_{\text{0,inter}}^2-\omega^2+i\gamma_{\text{inter}}\omega}\ ,
\end{equation}	
where $\varepsilon_\infty$, $\omega_{\text{Drude}}$, $\gamma_{\text{Drude}}$, $\omega_{\text{p,THz}}$, $\omega_{\text{0,THz}}$, $\gamma_{\text{THz}}$, $\omega_{\text{inter}}$, $\omega_{\text{0,inter}}$ and $\gamma_{\text{inter}}$ are treated as fitting parameters. We calculated the reflectance from transfer matrix method (see next section), and fitted the whole spectrum simultaneously. The fitted spectra agrees well with the experimental result as shown in Figure\,S2a. Along the direction that is perpendicular to the tube alignment direction, the dielectric constant is modeled as a frequency-dependent complex number $\varepsilon_{\perp}(\omega)$. We fitted the spectra point-by-point for each frequency. The following three points ensured that the recovered dielectric constants are unique and reliable: (i) the film thickness is not a fitting parameter and determined by atomic force microscopy. (ii) The Drude and Lorentz models used in the parallel direction connect real and imaginary parts through Kramers-Kronig relation. (iii) The dielectric constants in the perpendicular direction have negligible imaginary parts since the absorption is small.

\begin{figure}[!h]
	\begin{center}
		\includegraphics[scale=0.6]{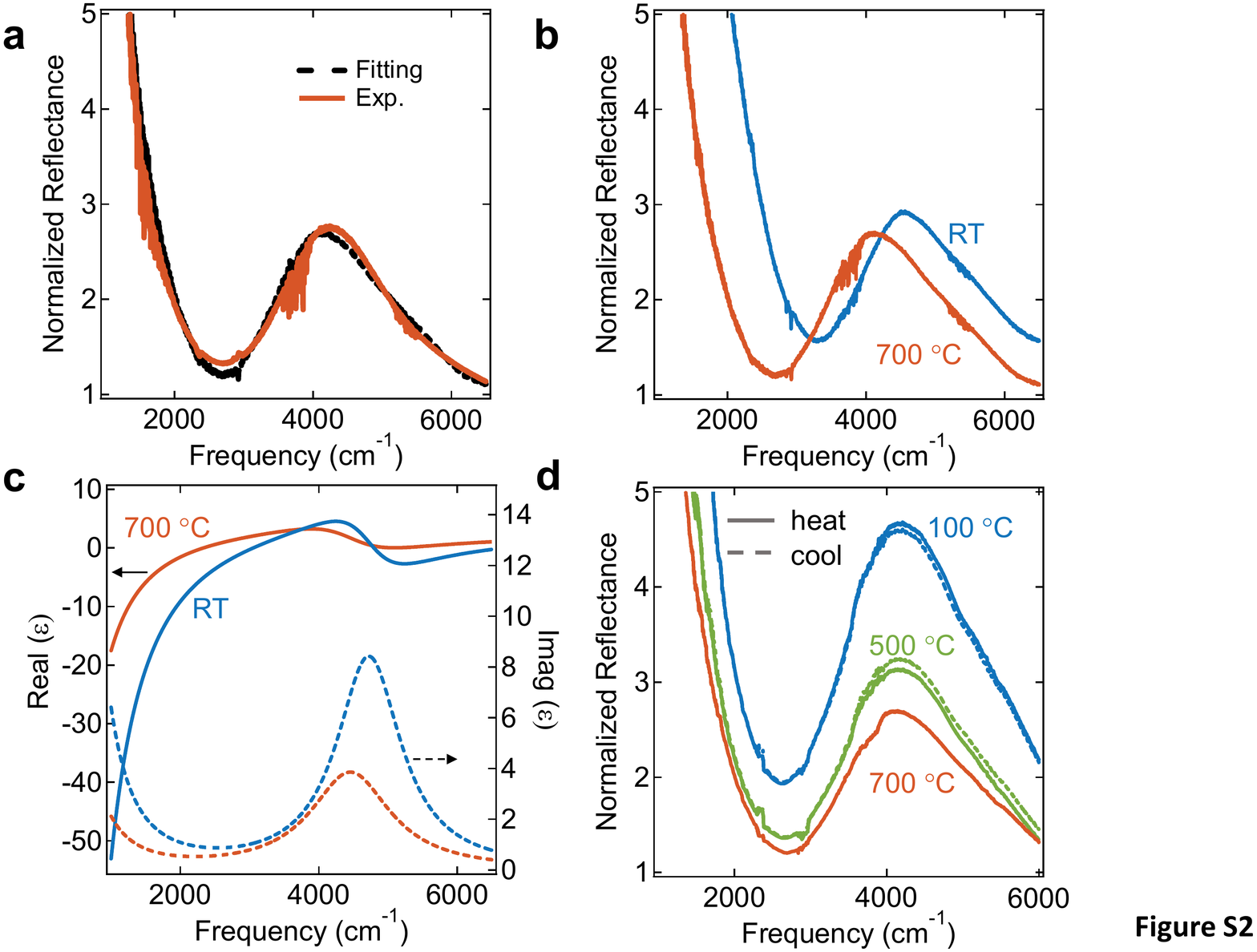}
		\caption{(a)~The experimental and fitting reflectance spectrum of a 110-nm thick film on a CaF$_2$ substrate at 700\,$\celsius$ when the detection polarization is along the nanotube alignment direction. (b)~Reflectance spectra of the film at room temperature (RT) and 700\,$\celsius$ for the first heating, and (c)~extracted dielectric constants. A strong dedoping effect is observed. (d)~Reflectance spectra of the film at various temperatures for more heating and cooling cycles.}
	\end{center}
\end{figure}

In addition, we compare the reflectance spectra and extracted dielectric constants along the tube alignment direction at room temperature and 700\,$\celsius$ for the first heating cycle. We clearly observe the spectra redshift in Figure\,S2b and the reduction of metallicity in Figure\,S2c at higher temperature. We attribute these behaviors to the SWCNT dedoping, as adsorbed water and gas molecules are removed from SWCNTs at 700\,$\celsius$ and under high vacuum. Once SWCNTs are dedoped, the optical properties of SWCNTS are very stable at high temperature and we do not see any noticeable change during the measurement, which usually lasted for tens of hours. Furthermore, we measured the reflectance spectra for more thermal cycles, and they agree well between the heating half-cycle and cooling half-cycle, as shown in Figure\,S2d. 

\subsection{Additional data on SWCNT hyperbolic cavities}
Figure\,S3 summarizes the mappings of the divergence of Poynting vector (Figures\,S3a--c and $E_\text{z}$ field distribution (Figures\,S3d--f) for (0.56,0.38), (0.63, 0.58) and (0.70, 1.05) cavities, respectively, and $E_\text{x}$ field distribution (Figures\,S3h and S3h) for (0.63, 0.58) and (0.70, 1.05) cavities, respectively. All three cavities shown in Figure\,3c of the main text support identical optical modes, and thermal emission is entirely from SWCNT cavities. We demonstrated hyperbolic cavities on the same isofrequency contour theoretically in Figure\,3c of the main text and experimentally in Figures\,4c and 4d, as we tuned $L_{\parallel}$ and $L_{\perp}$ simultaneously. 

\begin{figure}[!h]
	\begin{center}
		\includegraphics[scale=0.7]{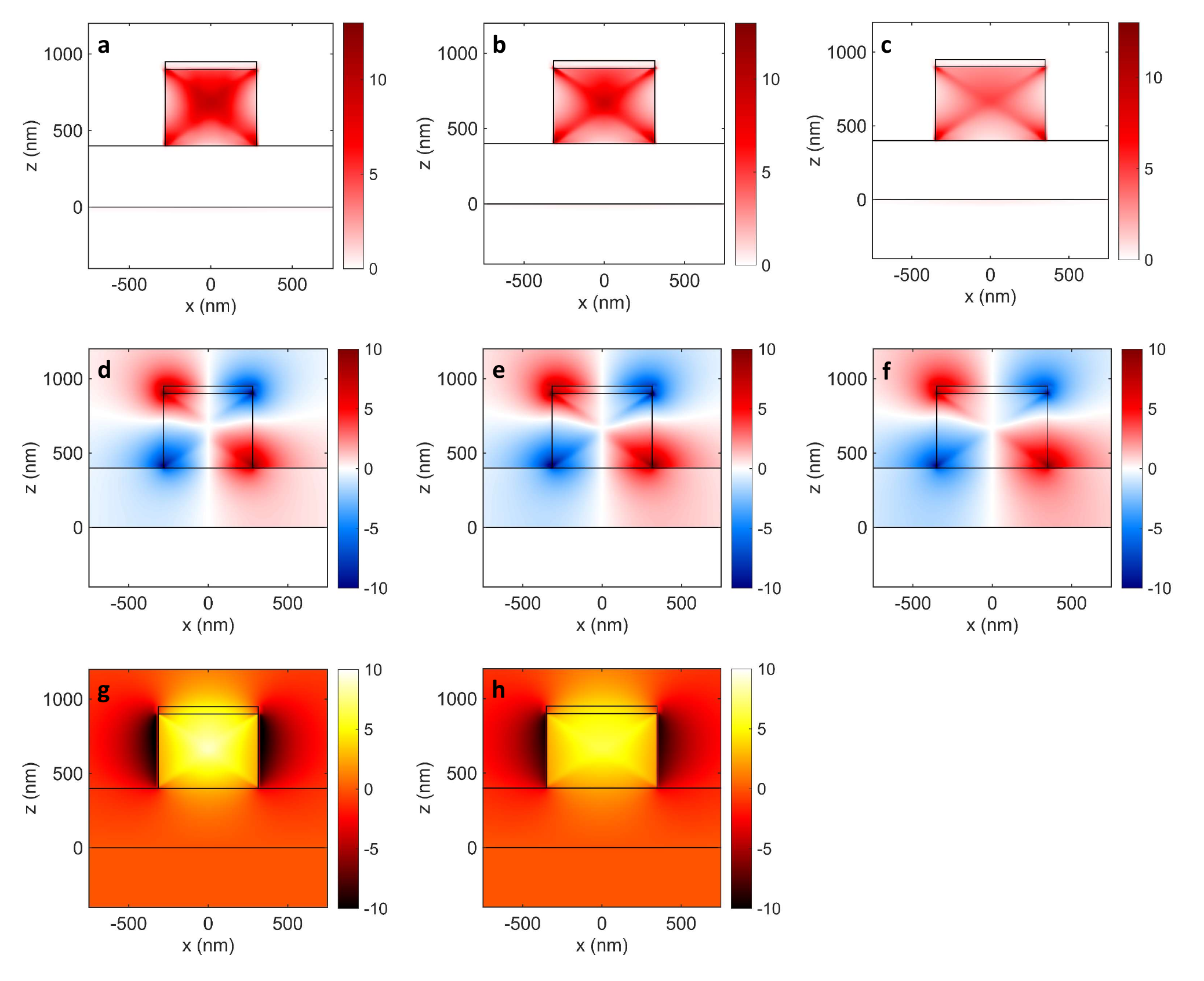}
		\caption{(a--c)~FDTD-calculated divergence of Poynting vector and (d--f),~$E_\text{z}$ electric field distribution for (0.56,0.38), (0.63, 0.58) and (0.70, 1.05) cavities, respectively. (g,\,h),~FDTD-calculated $E_\text{x}$ electric field distribution for (0.63, 0.58) and (0.70, 1.05) cavities. }
	\end{center}
\end{figure}

Figures\,S4 shows the calculated (Figure\,S4a) and experimental (Figure\,S4b) emission spectra for cavities with only one adjustable parameter $L_\perp$. In the hyperbolic region, the dielectric constant tensor of SWCNTs can be expressed as $diag(\varepsilon_\parallel,\varepsilon_\perp,\varepsilon_\perp)$, where $\varepsilon_\parallel<0$ and $\varepsilon_\perp>0$. The isofrequency contour is described as
\begin{equation}
\frac{k_x^2}{\varepsilon_\perp}+\frac{k_y^2+k_z^2}{\varepsilon_\parallel}=\bigg(\frac{\omega}{c}\bigg)^2\ ,
\end{equation}
where $k_x,k_y,k_z$ are wavevectors along $x,y,z$ axes. As $L_\parallel$ increases ($k_x$ decreases) with $\varepsilon_\perp>0$, we observe a normal scaling leading to a redshift ($\omega$ decreases in Fig.\,3b of the main text). On contrary, as $L_\perp$ increases ($k_y$ decreases and $k_z$ is fixed) with $\varepsilon_\parallel<0$, we observe an anomalous scaling leading to a blueshift ($\omega$ increases in Fig.\,S4). All these behaviors are consistent with the properties of hyperbolic cavities.

\begin{figure}[!h]
	\begin{center}
		\includegraphics[scale=0.6]{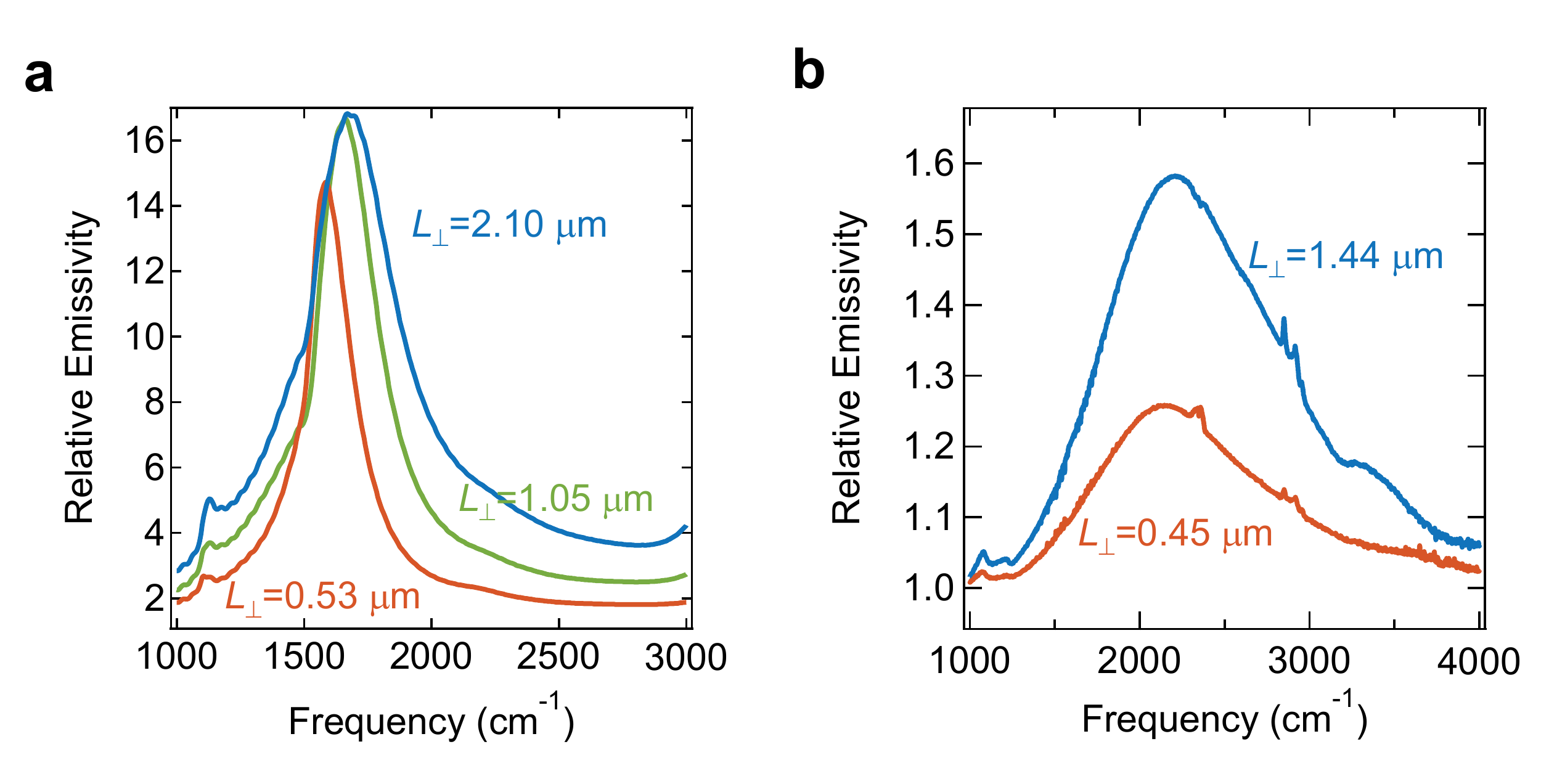}
		\caption{(a)~FDTD-calculated frequency-selective thermal emission from various cavities with tunable $L_{\perp}$ and fixed $L_{\parallel}=0.70\,\mu$m. (b)~Experimental frequency-selective thermal emission from various cavities with tunable $L_{\perp}$ and fixed $L_{\parallel}=0.60\,\mu$m.}
	\end{center}
\end{figure}
\medskip
\noindent {\em Acknowledgments.} W.\,G. and C.\,D. contributed equally to this work. W.\,G., X.\,L., J.\,K. acknowledge support by the Basic Energy Science (BES) program of the US Department of Energy through grant no. DE-FG02-06ER46308 (for preparation of aligned carbon nanotube films), the US National Science Foundation through grant no. ECCS-1708315 (for optical measurements), and the Robert A. Welch Foundation through grant no. C-1509 (for structural characterization measurements). 





\bibliography{jun,weilu}

\end{document}